# BIODEGRADABLE POLYLACTIC ACID (PLA) MICROSTRUCTURES FOR SCAFFOLD APPLICATIONS

*Gou-Jen Wang*[*], *Kuan-Hsuan Ho, and Cheng-Chih Hsueh*

Department of Mechanical Engineering,
National Chung-Hsing University,
Taichung, Taiwan
Email:gjwang@dragon.nchu.edu.tw

## ABSTRACT

In this research, we present a simple and cost effective soft lithographic process to fabricate PLA scaffolds for tissue engineering. In which, the negative photoresist JSR THB-120N was spun on a glass subtract followed by conventional UV lithographic processes to fabricate the master to cast the PDMS elastomeric mold. A thin poly(vinyl alcohol) (PVA) layer was used as a mode release such that the PLA scaffold can be easily peeled off. The PLA precursor solution was then cast onto the PDMS mold to form the PLA microstructures. After evaporating the solvent, the PLA microstructures can be easily peeled off from the PDMS mold. Experimental results show that the desired microvessels scaffold can be successfully transferred to the biodegradable polymer PLA.

## 1. INTRODUCTION

Biodegradable polymers are currently used in a number of biomedical applications, including sutures, screws, surgical adhesives, drug delivery, as well as the scaffolds for tissue engineering [1-2]. Polylactic acid (PLA) and its copolymers are commonly used biodegradable polymers in soft lithography [3-4] to fabricate scaffolds [5-7] due to their favorable biocompatibility and biodegradability. PLA is derived from lactic acid and can be easily produced in a high molecular weight form through ring-opening polymerization using a stannous octoate catalyst.

The soft lithography process includes fabrication of the elastomeric mode and use of it to make features in geometries defined by the mode's relief structure. Conventional soft lithography uses polydimethylsiloxane (PDMS) elastomeric mode formed from patterned silicon wafers or quarts to generate biodegradable microstructures by replica molding, hot embossing, and contact printing [8-9]. The main problem of conventional replica molding method is that the etching in silicon wafer is time consuming and can only fabricate relatively low aspect ratio devices. In addition, inclined side walls due to wet etching are inevitable, especially for high aspect ratio microstructures. Although negative photoresist SU-8 has been applied to fabricate master mode with high aspect ratio and near-vertical side walls [10-11], JSR photoresist has been found to be easily striped with no residues in solvent stripper solutions, making it suitable for the processing of MEMS devices [12].

In this research, we focus on developing a simple and cost effective process to fabricate PLA scaffolds for tissue engineering, using the negative photoresist JSR THB-120N as the master to cast an PDMS elastomeric mold, followed by spin-coating a thin poly(vinyl alcohol) (PVA) layer on it as a mode release such that the PLA scaffold can be easily peeled off. The PDMS mold is then filled with PLA precursor solution. After evaporating the solvent, the PLA microstructures can be peeled off from the PDMS mold and be used for cell culture.

## 2. MATERIALS AND METHODS

Soft lithography was implemented to fabricate PLA microstructures for tissue engineering by dissolving PLA in Dioxane (TEDIA) then casting the solution onto the elastomeric PDMS mold. After evaporating the solvent, PLA layer was stripped off from the replica mode and microstructures were transferred from the replica mode to the PLA layer.

### 2.1 PLA polymer solution

PLA particles and dioxane (TEDIA) with ratio1:1 were used to prepare PLA solution. A magneto agitator was implemented to enable the PLA particles to completely be dissolved into the dioxane at 60 °C.

## 2.2 Master and elastomeric mold fabrications

JSR THB-120N spun on a glass substrate was photo-patterned through standard UV lithography and developed to be used as a master. There are two reasons to implement the JSR THB-120N negative photoresist. Firstly, it can be spin-coated to a relative thick film (10-500um in single coating) to fulfill the high aspect ratio requirement of the microstructures. Furthermore, it can be easily striped with no residues in solvent stripper solutions. A 60 μm thick JSR THB-120N layer was obtained by first spin-coating at 350 rpm for 10 sec followed by 500 rpm for 25 sec. The developer solution was THB-D1 by JSR Inc. Ideally the JSR THB-120N master can be directly used as the elastomeric mode. However, the JSR THB-120N master can also be easily dissolved by the solvent that is used to dissolve the PLA. It is thus desired to use PDMS that can resist the dissolution of the solvent as the elastomeric mold to further replicate the PLA microstructures.

Figure 1 schematically illustrates the manufacturing procedures of the PDMS elastomeric mold. The PDMS solution is cast onto the photoresist master that is contained in a vessel and placed at 70 °C for couple hours. After solidification, the PDMS mold was peeled off from the master for micro molding the PLA microstructures.

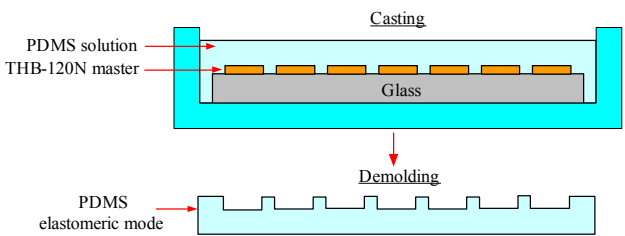

Figure 1. Fabrication procedures of the PDMS lastomeric mold

## 2.3 Scaffold fabrication

The procedures to fabricate the PLA scaffold are illustrated in Figure 2. PVA (1%) solution was spun on the PDMS mold as the mode release such that the PLA scaffold can be easily peeled off from the mold. The spin-coating parameters for a 6 μm thick PVA film were 1000 rpm for 25 sec followed by 1500 rpm for 10 sec. Since the thickness of the PVA layer affects the dimensions of the final product, dimensions of the PDMS mold should include the effect of the PVA layer.

The PLA solution was cast onto the PDMS mold that was contained in a vessel and placed at room temperature for 180 min to evaporate the dioxane. Immerse the vessel into DI water for couple hours to hydrolyze the PVA film and peel the PLA microstructure from the PDMS mold.

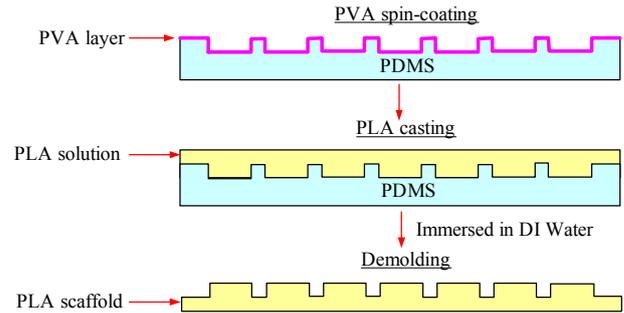

Figure 2. Fabrication procedures of the PLA scaffold

## 3. EXPERIMENTAL RESULTS AND DISCUSSIONS

Figure 3 shows a CAD drawing of a microvessel scaffold for tissue engineering that was the object to be fabricated; in which, the diameter with respect to each microchannel ranges from 60μm to 120μm and the length of each segment is 800μm. One of the major challenges of the tissue engineering is the lack of intrinsic blood vessels to transport nutrient and metabolite. Once the tissue is larger then 1-2 mm, the cultivating cells will shrivel due to the lack of metabolic nutrient and air. It is thus desired to provide the artificial tissues with artificial microvessels. To grow artificial microvessels, a micro scale scaffold to cultivate endothelial cell has to be fabricate. Since the microvessels are to be implanted into a living body, the material of the scaffold needs to be biodegradable.

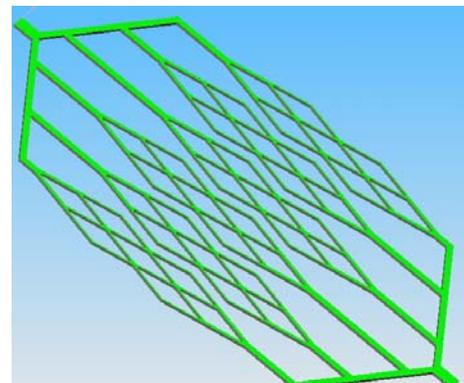

Figure 3. Microvessel scaffold for tissue engineering

Figure 4 illustrates the OM images of JSR THB-120N convex master. It can be found that the desired microchannel structures are completely transferred to the photoresist.

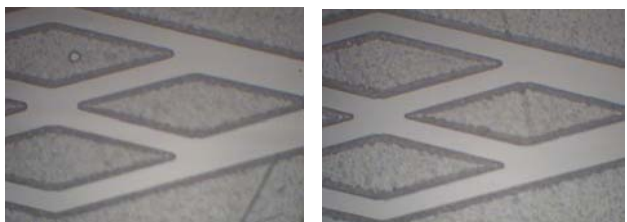

Figure 4. OM images of the JSR THB-120N convex master

The OM images of the PDMS concave mold are presented in Figure 5. The microstructures are successfully replicated onto the PDMS from the photoresist master.

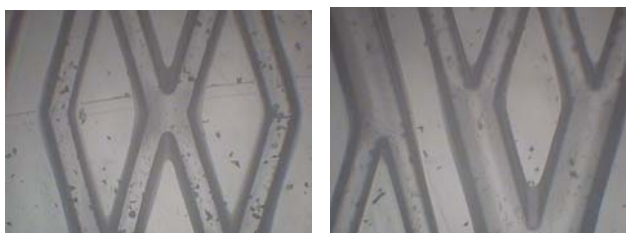

Figure 5. OM images of the PDMS concave mold

The MO micrographs of the PLA microstructures peeled off from the PDMS mold is depicted in Figure 6. The microstructures can be used to seed cells. Bonding the PLA microstructures with a flat PLA plate, a PLA scaffold for microvessels can be obtained.

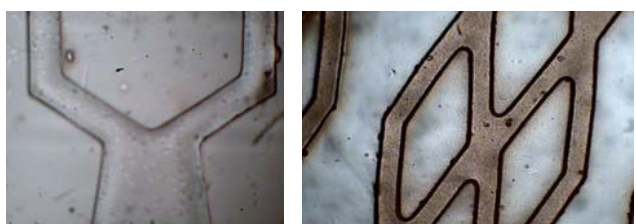

Figure 6. OM micrographs of the PLA scaffold

### 4. CELL SEEDING

The seeding cells used to grow the microvessels are the bovine endothelial cells (BEC). Semi-dynamic seeding is executed. Schematic illustration of the semi-dynamic seeding is shown in Figure 7. Instead of using a peristaltic pump, circulation of the cultivation medium is carried out by periodically injecting fresh medium into and sucking aging medium out of the scaffold. The circulation frequency is 2-3 times daily.

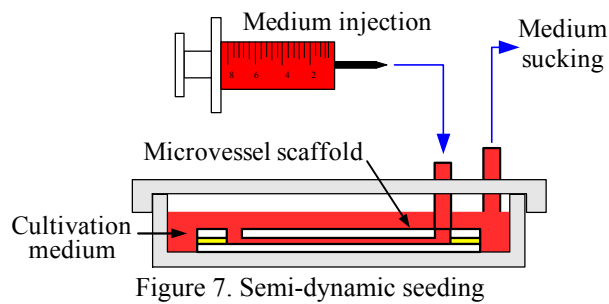

Figure 7. Semi-dynamic seeding

### 4.1 Biocompatibility Experiments

Biocompatibility is the key factor that determines the adhesion between cells and the scaffolds. Different kinds of biomaterials, TCPS (conventional used biomaterial for cell seeding), PDMS, PLA, and glass substrate are used for biocompatibility comparisons. Due to its high hydrophilic identity, the glass scaffold is used for contrast. Biocompatibility experimental results are shown in Figure 8. The TCPS scaffold has the largest number of living cells, followed by the glass, PLA and PDMS. It is demonstrated that the PLA has better biocompatibility than the PDMS scaffold.

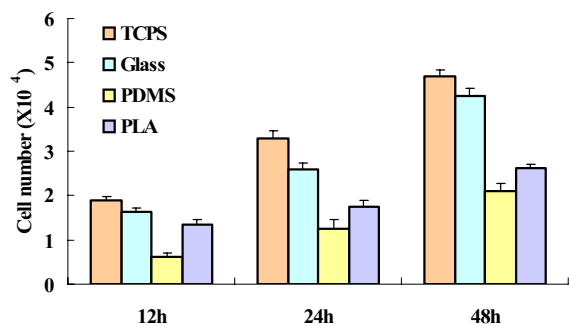

Figure 8. Biocompatibility experiments

### 4.2 Semi-dynamic seeding

Figure 9 shows the seeding processes of the cell suspension with a concentration of $10^6$ cells/ml. After 72 hours of seeding, it is observed that cells completely adhere to the microchannel toward near-confluence.

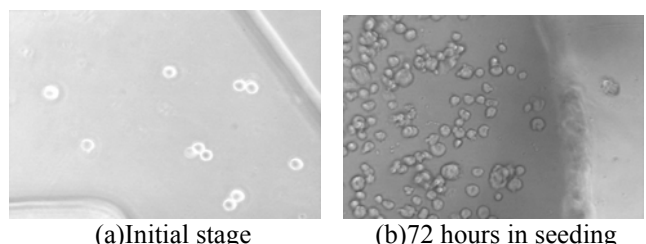

(a) Initial stage    (b) 72 hours in seeding

Figure 9. Cells seeding for the $4\times10^6$ cells/ml cell suspension

## 5. CONCLUSIONS

In this article, a simple and cost effective soft lithographic process to fabricate PLA scaffolds for tissue engineering was presented. The negative photoresist THB-120N by JSR Inc. that can be easily striped with no residues in solvent stripper solutions was implemented to fabricate was implemented to fabricate the master to cast the PDMS elastomeric mold. A thin PVA layer was then spun on the PDMS mold to get easy release of the PLA microstructures. The PLA precursor solution that was prepared by dissolving PLA particles into dioxane with ratio1:1 was cast onto the PDMS mold. After solidification, the PLA microstructures can be peeled off from the PDMS mode and can be future used to cultivate cells. Our future work will focus on fabricating different nano patterns on the scaffold surface to examine the inferences of nono features on cell growth.

## ACKNOWLEDGEMENTS

The authors would like to thank the National Science Council of Taiwan, for financially supporting this work under Contract No. NSC-94-2212-E-005-010. The Center of Nanoscience and Nanotechnology at National Chung-Hsing University, Taiwan, is appreciated for use of its facilities.